\documentclass[useAMS,usenatbib]{mn2e}

\usepackage{graphicx}
\usepackage{float}
\usepackage{xcolor}
\usepackage{aecompl} 
\usepackage[T1]{fontenc}

\newcommand{\msun}{~M$_\odot$~}

\title[Li in POP II stars from PMS to Spite plateau]
{Lithium evolution in metal-poor stars: from Pre-Main Sequence to the Spite plateau}
\author[X. Fu, et al]{Xiaoting Fu$^{1}$\thanks{E-mail:
xtfu@sissa.it}, Alessandro Bressan$^{1}$, Paolo Molaro$^{2}$, Paola Marigo$^{3}$
\\
$^{1}$SISSA - International School for Advanced Studies, via Bonomea 265, 34136 Trieste, Italy\\
$^{2}$INAF - Osservatorio Astronomico di Trieste, via G. B. Tiepolo 11, 34143 Trieste, Italy \\
$^{3}$Dipartimento di Fisica e Astronomia, Universit\`{a} di Padova, Vicolo dell'Osservatorio 2, I-35122 Padova, Italy}

\begin{document}

\date{Accepted 2015 June 18.  Received 2015 June 18; in original form 2015 March 11}

%\pagerange{\pageref{firstpage}--\pageref{lastpage}} \pubyear{2002}

\maketitle

%\label{firstpage}

\begin{abstract}
Lithium abundance derived in metal-poor main sequence stars is about three times
lower than the value of primordial Li predicted by the standard Big Bang
nucleosynthesis when the baryon density is  taken from the CMB or the deuterium
measurements.  This disagreement is generally referred as the lithium problem.  We
here reconsider the stellar Li evolution from the pre-main sequence  to the end
of the main sequence phase by  introducing  the effects of convective
overshooting and residual mass accretion.  We show that   $^7$Li  could be
significantly depleted by convective overshooting in the pre-main sequence phase
and then partially restored in the stellar atmosphere by a tail of matter
accretion which follows the Li depletion phase and that could be
regulated by EUV photo-evaporation.
By considering the conventional nuclear burning and  microscopic diffusion
along the main sequence we can  reproduce the Spite plateau for stars
with initial mass $m_0=0.62~-~0.80~M_{\odot}$, and the Li declining
branch  for  lower mass dwarfs, e.g, $m_0=0.57~-~0.60~M_{\odot}$, 
for a wide range of metallicities
(Z=0.00001 to Z=0.0005), 
starting from an initial  Li abundance $A({\rm Li}) =2.72$.
This environmental Li evolution model also offers the
possibility to interpret the decrease of Li abundance in extremely metal-poor
stars, the Li disparities in spectroscopic binaries and the low Li abundance in
planet hosting stars.

\end{abstract}


\section{Introduction}

${}^7$Li is one of the four isotopes synthesized in the primordial nucleosynthesis.
Its primordial abundance depends mainly on the baryon-to-photon ratio
with only a minor sensitivity to the universal speed-up expansion rate,
namely the number of neutrino families \citep{Fields11,fms2014}.
The universal baryon density can be obtained
either from the acoustic oscillations of the cosmic microwave background (CMB) observation
or independently,
from the primordial deuterium abundance measured in un-evolved clouds of distant quasar spectra \citep{adams1976}.
Observations based on the Wilkinson Microwave Anisotropy Probe \citep[WMAP;][]{wmap}
predict a primordial ${}^7$Li abundance
$A({\rm Li}) =2.72$\footnote{$A({\rm Li}) = 12 + \log [n({\rm Li})/n({\rm H})]$
where $n$ is number density of atoms and 12 is the solar hydrogen abundance.}
\citep{coc2012}.
From the baryon density measured by the
Planck mission \citep{planck} \citet{coc2014}
calculated the primordial value of ${}^7$Li/H
to be $4.56\sim5.34\times10^{-10}$~(A(Li)$\approx2.66-2.73$).

POP II main sequence (MS) stars show a constant ${}^7$Li abundance~\citep{spite1982},
which was interpreted as an evidence that the these
stars carry the primordial $^7$Li abundance because of their low metallicity.
In the past three decades,
observations of metal-poor main sequence stars both in the Milky Way halo
\citep{spite1982, Sbordone}, and in the globular clusters \citep{lind09}~
confirm that the ${}^7$Li abundance remains $A({\rm Li}) \approx 2.26$ \citep{molaro08}.
This abundance, which defines the so-called Spite plateau, is three times lower
than the predicted primordial value.  The  discrepancy is
the long-standing ~\textquotedblleft lithium problem\textquotedblright.

There are several lines of study which have been pursued to provide possible
solutions to the problem:
i) nuclear physics solutions which
alter the reaction flow into and out of mass-7 \citep{coc2012}; 
ii) new particle physics where massive decaying particles could destroy $^7$Li
\citep{olive12, kajino12};
iii) Chemical separation by magnetic field in the early structure formation 
that reduces the abundance ratio of Li/H \citep{kusakabe};  
iv) $^7$Li depletion during main sequence evolution.
It has been argued that certain physical processes,
which may occur as the stars evolve on the main sequence, 
could cause the observed lithium depletion.
Among these processes we recall 
gravitational settling (e.g. \citet{sw2001,richard2005} and \citet{korn2006}) 
or possible coupling between internal gravity waves and rotation-induced mixing \citep{ch05}.

In this paper we address the lithium problem 
on the ground of the Pre-Main Sequence (PMS) stellar physics,
under the assumption that the standard BBN is correct.
Hereafter when not otherwise specified for Li we refer to the isotope 7.
The surface Li evolution during the PMS has been analyzed
in stellar evolution models with solar metallicity by several authors
\citep{dantona1984,soderblom93,dantona1994,swenson94,ventura98, piau02,
Tognelli12}, 
with the general conclusion that mixing efficiency plays a key role on Li depletion.
Recent extensive studies in metal-rich young open clusters do show
that the observed lithium abundance is modified during the PMS phase ~\citep{somers2014}.
In this respect, two points are worthy of consideration.
On one side,  it has been shown that overshoot at the base
of the solar convective envelope is
more efficient than hitherto believed ~\citep{overshoot},
and could be even more in the envelope of a PMS star.
This would favour a more efficient Li depletion.
On the other side,
there is evidence that a residual mass accretion (also called late accretion) persists
for several $10^7$~years and is observed up to the early main sequence phase
\citep{demarchi2011}.
This may act to partially restore Li in metal-poor PMS stars.
\citet{molaro12}~first proposed that Li abundance both in the Population I (POP I)
 and Population II (POP II) stars
could be modified due to the combined effects of efficient overshoot
and late mass accretion during the PMS evolution.
In this study we use \texttt{PARSEC}
\citep[PAdova and TRieste Stellar Evolution Code;][]{parsec}
to examine the working scenario proposed by \citet{molaro12} in a more systematic way.

The structure of the paper is as follows.
Section~2 describes the theoretical pre-main sequence models, which
include envelope overshooting, residual accretion and EUV photo-evaporation.
Section~3 describes the Li evolution in main sequence.
Section~4 presents the results of the models and the comparison with the observations.
An ample discussion and the main conclusions are drawn in Section~5.

\section{Pre-Main Sequence Li evolution}\label{sec:pms}

Pre-main sequence is the direct continuation of the proto-stellar phase.
Initially, as the young stellar object (YSO) evolves along its stellar birthline,
its luminosity is mainly supported by an accretion process strong enough
to maintain active deuterium fusion \citep{stahler83}.
Once the accretion ceases, the proto-star, surrounded only by a residual disk,
descends along its Hayashi line almost vertically
in the Hertzsprung-Russell diagram (H-R diagram),
undergoing a rapid gravitational contraction.

In this phase evolution of PMS stars critically depends on their mass,
and lithium can be burned at different stages
through the reaction $^7$Li(p,$\alpha)^4$He.
Lithium is very fragile as
the nuclear reaction rate $R_{nuclear}$~of ${}^7$Li(p,$\alpha)^4$He
becomes efficient already at temperatures of a few million Kelvin.
The effective Li burning temperature for PMS stars,
i.e. that needed to consume Li in a timescale of $\sim~10^7~yr$,
is $\sim~4\times10^6~K$.
The $R_{nuclear}$~is adopted from JINA REACLIB database~\citep{jina}.

Very low mass stars with initial mass $m_0 <0.06 M_{\odot}$ never reach
this temperature. More massive PMS stars experience Li-burning,
which initially affects the entire stellar structure as long as it is fully convective.
Later, at the formation of the radiative core, the extent of Li
burning can vary, depending on the mass of the star.
Thereafter, the  Li evolution is critically
affected the temperature at the base of the convective envelope,
and the efficiency of overshoot at the base of the convective envelope
begins to play a significant role.

We use the stellar evolution code \texttt{PARSEC}
to calculate the PMS evolution of low mass stars with initial mass from $0.50~M_\odot$~to~$0.85~M_\odot$,
with metallicity $Z=0.0001~(\sim [M/H]=-2.2$ dex), typical of POP II stars.
The adopted helium abundance is $Y_p =0.249$, based on the helium enrichment law \citep{parsec}:
\begin{equation}
 Y = Y_p +\frac{\bigtriangleup Y}{\bigtriangleup Z}Z = 0.2485 + 1.78Z
\end{equation}
where
$Y_p=0.2485$~is the adopted primordial value 
and $\bigtriangleup Y / \bigtriangleup Z$~is the helium-to-metals enrichment ratio.
Following the SBBN prediction, we set the initial lithium abundance A(Li)=2.72.
A solar-model calibrated mixing length parameter~$\alpha_{MLT}=1.74$~is adopted.

\subsection{Convective overshooting}\label{sec:ov}

Overshooting (OV) is the signature of the non-local convective mixing in the star
that may occur at the borders of any convectively unstable region \citep{bressan14}.
As already anticipated,
efficient overshooting at the base of the convective envelope may significantly affect PMS surface Li depletion,
because overshooting extends the mixed region into the hotter stellar interior.

Recently~\citet{overshoot} suggest that an overshooting region of
size $\Lambda_e \approx~ 0.3 \sim 0.5~H_p$
(where $H_p$~is the pressure scale height),
at the base of the convective envelope of the Sun,
provides a better agreement with the helioseismology data.
These values are higher than those of earlier studies \citep[e.g.,][]{miglio2005},
and indicate that the overshoot mixing could be much more efficient than previously assumed.
It should be recalled that by applying an efficient envelope overshoot,
many other observational data could be better reproduced,
such as the blue loop of intermediate and massive stars~\citep{alongi},
the low-mass metal-poor stars~\citep{parsec},
the carbon stars luminosity functions in the Magellanic Clouds via a more
efficient third dredge-up in AGB stars~\citep{herwig00, MarigoGirardi_07},
and the location of the RGB bump in old star clusters~\citep{alongi}.

\begin{figure}
 \centering
 \includegraphics[width= .45\textwidth,angle=0]{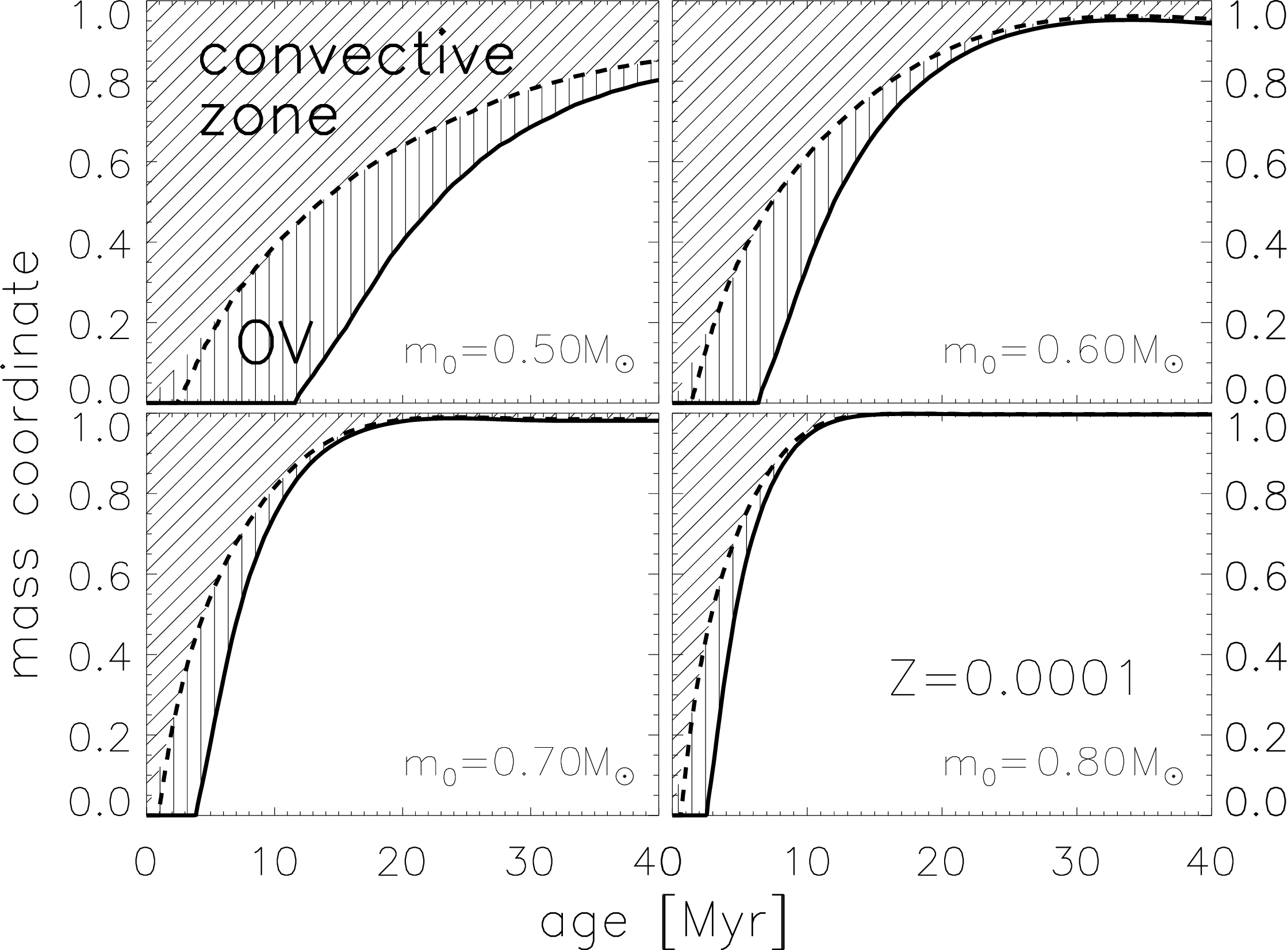}
 \caption{
 Kippenhahn diagrams for stars with initial metallicity Z=0.0001  and  different initial masses.
 The hatched area corresponds the mass size of the convective zone,
 within it the area filled with vertical lines
represents the contribution of the envelope overshooting.
 On the y-axis the mass coordinate of the convective zone starts from 0
(at the center of the star) which means the star is fully convective,
 while 1 means the base of the convective zone is at the surface of the star
and there is no convection.
 As the star evolves, the base of the convective zone retreats towards the surface of the star,
 and the overshoot layer become shallower and shallower.
}
 \label{fig:conv}
\end{figure}

Soon after the formation of the radiative core we assume that the
overshooting at the base of the deep convective envelope is quite efficient,
with~$\Lambda_e\sim~1.5~H_p$.
This value is intermediate between the one suggested for the Sun \citep{overshoot}
and the larger values ($\Lambda_e=2-4~H_p$) suggested by  \citet{tang} for intermediate and massive stars.
We will show in Sect.~\ref{sec:dis} that this value is not critical, since
even assuming a maximum overshoot $\Lambda_e\sim~0.7~H_p$,
the model can well reproduce the data.
The overshooting distance is computed using the natural logarithm of the pressure,
downward from the Schwarzschild border (P$_{sch}$) at the
bottom of the convective envelope, i.e. an underlying stable layer (P$_{lay}$) is mixed with the
envelope if ln(P$_{lay}$)~-~ln(P$_{sch})\leq\Lambda_e$.
Then, as the star moves toward the zero age main sequence (ZAMS),
we vary the overshooting efficiency proportionally to the mass size
of the outer unstable convective region~($f_{cz}$),
until the typical value estimated for the Sun~$\Lambda_e=0.3~H_p$~\citep{overshoot}
is reached:
\begin{equation}
 \Lambda_e = 0.3 + (1.5 - 0.3)* f_{cz}
 \label{eq:ov}
\end{equation}
This assumption produces an efficient
photospheric Li depletion during the PMS phase, while preserving the solar constraints for the main sequence stars.
 Fig.~\ref{fig:conv} illustrates the temporal evolution of the size
of the convective and overshoot regions for four values of the initial stellar mass,
$m_0=0.5,0.6,0.7,0.8~M_\odot$~at metallicity Z=0.0001.
The mass coordinate, on the Y axis, goes from 0, at the center of the star,
to 1 at the surface.
The stars are initially fully convective
and the temperature at the base of the convective zone~($T_{bcz}$)~is high enough to burn Li so efficiently to begin surface Li depletion.
As the central region heats up, the base of the convective envelope moves outwards to the stellar surface, while~$T_{bcz}$~decreases.
When $T_{bcz}$~drops below the lithium effective burning threshold,
the photospheric Li-depletion ceases,  leading to an almost total or only partial Li depletion,
depending on the stellar mass and on the overshooting efficiency.

 \begin{figure}
 \centering
 \includegraphics[width= .45\textwidth,angle=0]{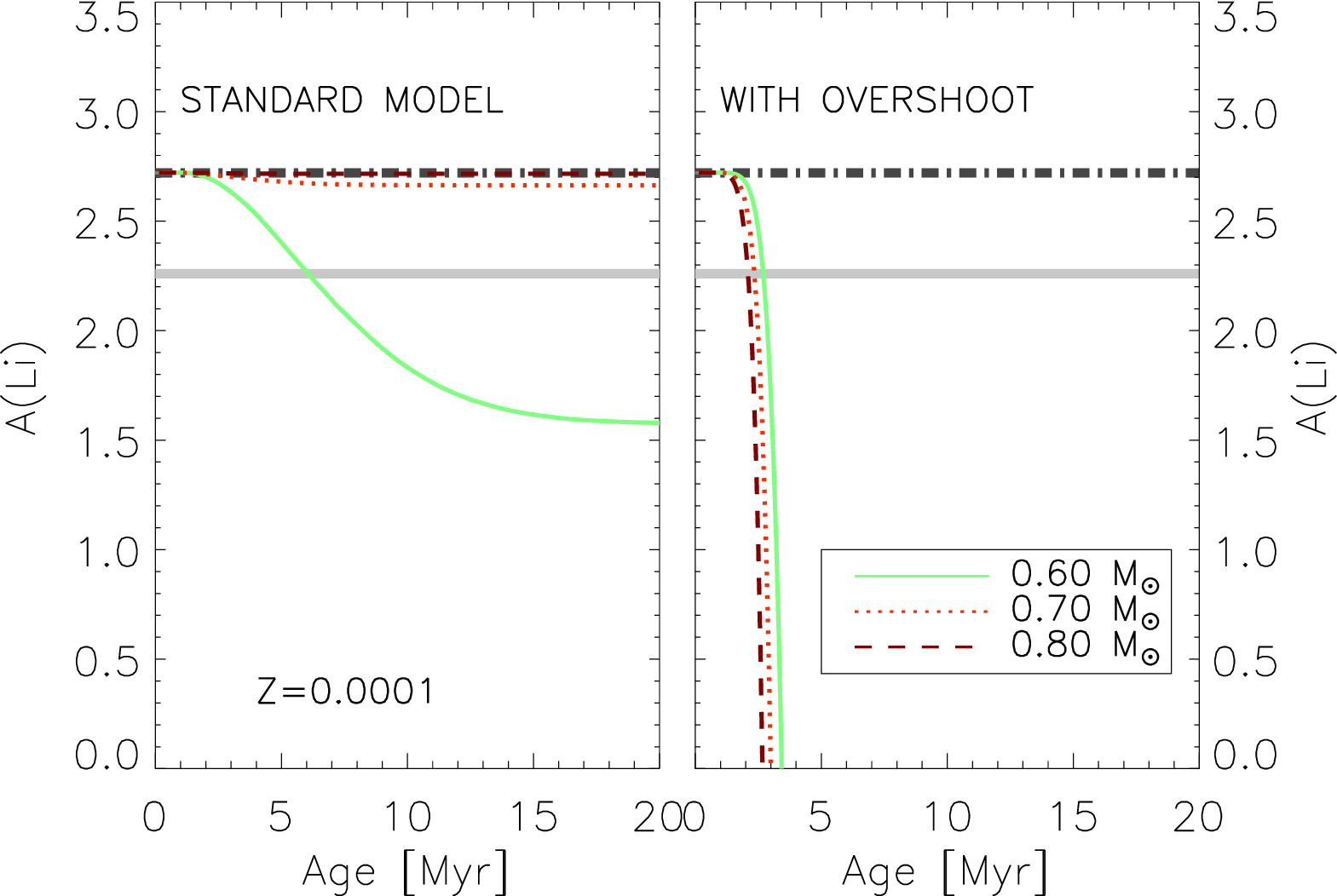}
 \caption{PMS Li evolution in the standard model (left panel) and model
with envelope overshoot (right panel) for three stellar models under consideration.
 The two horizontal lines indicate primordial A(Li) (dark grey dot dashed line)
 and Spite plateau A(Li) (light grey solid line), respectively.
Overshooting is the solo parameter tested, no other mechanism is applied.}
 \label{fig:pmsov}
\end{figure}

Figure~\ref{fig:pmsov} illustrates the effect of efficient overshooting as the
\textit{solo} mechanism applied to the PMS phase for $m_0=0.60,~0.70,~0.80~M_\odot$ stars.
In standard model without overshoot (left panel), A(Li) almost remains
the same as the primordial value, unless the convective zone in the star itself
goes  deep enough to burn Li,  as for $m_0=0.60~M_\odot$ star in the left panel.
Conversely, assuming efficient envelope overshooting,
Li abundance is found to be fully depleted over a few Myr in the same stellar models
(right panel).
It follows that if overshoot were the only process at  work,
a very low Li abundance should be measured at the surface of those stars.

\begin{figure}
 \centering
 \includegraphics[width=.45\textwidth,angle=0]{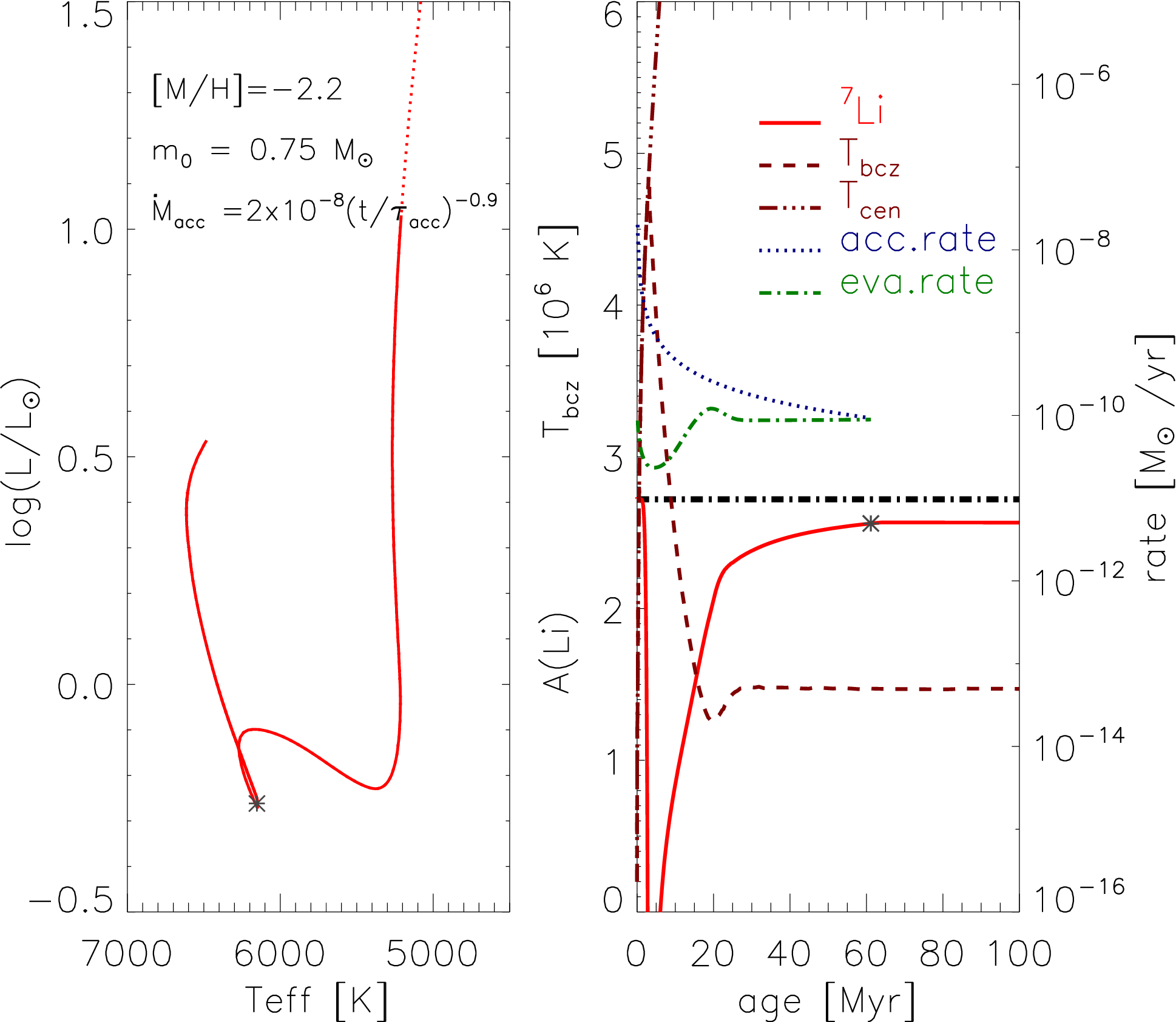}
 \caption{An example of the PMS lithium evolution for a star with $m_0 = 0.75 M_{\odot}$, [M/H]=-2.2.
 Left panel: Evolutionary track at constant mass in H-R diagram.
 The solid line starts at the end of the dotted-line-indicated deuterium burning phase. 
 At this stage the accretion is reduced to $2\times10^{-8}~M_{\odot}/$yr.
 The asterisk marks the end of the residual accretion.
 Right panel: Li evolution starting from an initial abundance A(Li)=2.72 dex (horizontal black dot dashed line).
 The temperatures at the center ($T_{cen}$, dark red dot dot dashed line) 
 and at the base of the convection zone ($T_{bcz}$, dark red dashed line) 
 are also shown.
 The accretion (with rate drawn by the dark blue dotted line) is terminated
 by the EUV photo-evaporation (with rate drawn by the dark green dot dashed line)
 when the latter reaches the same value as the former.
  }
 \label{fig:m75}
\end{figure}

\subsection{Late mass accretion}\label{sec:acc}
Recent observations of late PMS stars --
those that have already abandoned the stellar birth line and are joining the ZAMS --
reveal that most of them show H$_\alpha$-line emission,
which likely originates in a residual accretion process \citep{demarchi2010, spezzi2012}.
This suggests that the disk accretion may last much longer than previously believed:
at least tens of Myr and even up
to the early main sequence stars \citep{demarchi2011}.

This late accretion (also known as residual accretion) is different from the one that
maintains the star on the stellar birth line during the proto-stellar phase.
The observed values indicate a median accretion rate of ($\sim10^{-8}M_{\odot}$~/yr)
for YSO T Tauri stars (TTS) with a steady decline as time proceeds \citep{Espaillat2014}.
Most residual disks presented in~\citet{Espaillat2014}~and \citet{Gallardo2012} are associated with accretion rates
at the level of $\sim10^{-10}M_{\odot}/$yr - $10^{-8}M_{\odot}/$yr.

Since the accreting material keeps the initial Li abundance,
its contribution cannot be neglected, especially if the photospheric value
has been already depleted by another process.
In this framework we  modify our stellar evolution code \texttt{PARSEC}
to account for the effect of such a residual accretion during the PMS phase.

We assume that, 
after the main accretion phase when the star leaves the stellar birth-line
by consuming its internal deuterium,
a residual accretion keeps going on.
Since our PMS models are initially evolved at constant mass
from a contracting configuration without nuclear burning, 
we assume that the residual accretion  begins 
when deuterium burning ends, that is when the
photospheric deuterium abundance drops to 1/10 of its initial value, 
and indicate this time as $\tau_{acc}$.
We then apply the accretion to the models, starting at t=$\tau_{acc}$
and assume that it declines with time following a power law
  \begin{equation}
   \dot{M} = \dot{M_0} \left(\frac{t}{\tau_{acc}}\right)^{-\eta}\,\,\,\,\,\,\,\,[M_{\odot}/{\rm yr}]
  \label{equ:acc}
  \end{equation}
where $\dot{M_0}$ is the initial rate and $\eta$ is the parameter that specify the rate of decline.
In principle, the parameters  $\dot{M_0}$, $\tau_{acc}$, and $\eta$ could be treated
as adjustable parameters to be constrained with the observed rates,
after properly considering a detailed PMS evolution that includes the main accretion phase.
However, since in this paper we perform an explorative analysis
without a full description of the protostar phase
and the initial evolution with large accretion rates,
we assume the residual accretion begins when deuterium burning ends as defined before. 

Table~\ref{tab:tstart} shows the age at the beginning of the residual accretion
($\tau_{acc}$) for different stellar masses.
We note that the duration of  the early large accretion,
before $\tau_{acc}$, is around~$1-2~\times10^5$ yr, hence negligible
compared to the PMS lifetime (several~$10^7$~yr).
As to the initial accretion rate  $\dot{M_0}$, we assume
$\dot{M_0} = 2\times{10^{-8}} ~M_{\odot}/$yr,
which is a reasonable value close to the observed median
accretion rate for young TTS \citep{hartmann1998}.
The exponent $\eta$ describes how fast the accretion rate declines with increasing age $t$.
Here we set $\eta = 0.9$ to recover most of the observed
rates from~\citet{Espaillat2014,Gallardo2012,demarchi2011}.

 We also assume that during this residual accretion phase
the original material is accreted following Equation ~\ref{equ:acc},
irrespective of the geometry of the disk.

The right panel of Fig.~\ref{fig:m75} illustrates the effect of late accretion on the Li evolution,
for a star with initial mass $m_0=0.75M_{\odot}$.
The accretion material falling onto the star contains lithium with the initial abundance
and, even if accretion is very small after tens of million years,
it restores the surface  $^7$Li towards the initial value.

\subsection{EUV photo-evaporation}\label{sec:euv}
\begin{figure}
 \centering
 \includegraphics[width= .48\textwidth,angle=0]{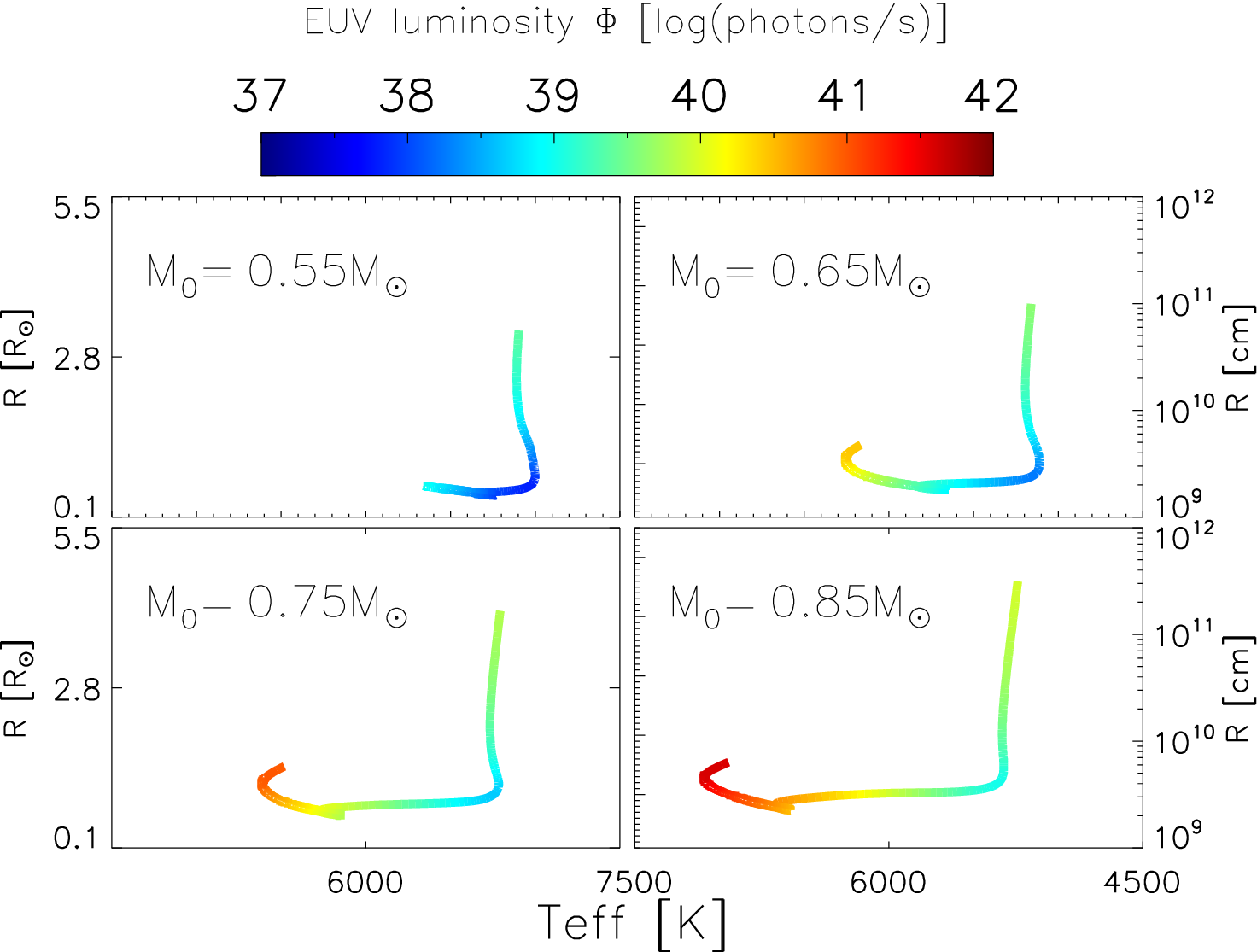}
 \caption{
Evolution of the EUV photon luminosity from the central stars (color-coded
according to logarithmic scale shown on the top bar)
as a function of the effective temperature and radius, assuming black-body emission.
Note the different units for the radius on the left and right Y-axes.
}
 \label{fig:euv}
\end{figure}

Late accretion will last until the remaining gas reservoir is consumed or until some feedback mechanism from the
star itself is able to clean the nearby disk.
In this respect, Extremely UV (EUV) radiation ($13.6 - 100$~eV, $10 - 121$~nm) photons could be
particular important in determining the end of the accretion process because
they have energy high enough to heat the disk surface gas.
The warm gas can escape from the gravitational potential of the central star
and flow away in a wind~\citep{Dullemond2007}.
Although no actual evaporative flow observation is available yet,
there is a large consensus that this radiation could significantly reduce the disk mass and shorten the lifetime of the disk.
The existence of pre-transitional and transitional disks with a gap or a hole in the disk
are taken as possible evidence of this process~\citep{Dullemond2007,Espaillat2014}.
The magnitude of the disk mass loss caused by EUV evaporation is given by the following relation~\citep{Dullemond2007}:
\begin{equation}
\dot{M}_{EUV} \sim 4\times10^{-10} \left(\frac {\Phi_{EUV}} {10^{41}s^{-1}}\right)^{0.5}
\,\left(\frac{M_{\ast}}{M_{\odot}}\right)^{0.5}\,\,[M_{\odot} {\rm yr}^{-1}]
\label{equ:euv}
\end{equation}
where $\Phi_{EUV}$ is the EUV photon luminosity [photons/s] produced by the central star.
This effect can be easily included in \texttt{PARSEC} under the assumption that the stars
emit as a black body at the given effective temperature $T_{\rm eff}$
In Fig.~\ref{fig:euv} we show the EUV luminosity evolution for stellar models
with initial masses $m_0= 0.55, 0.65, 0.75,$ and $0.85\, M_{\odot}$.
The photon luminosity $\Phi_{EUV}$ varies with the effective temperature and
the radius of the central star, therefore being importantly affected by
the stellar evolutionary phase.
Most of the EUV luminosities range from $\Phi_{EUV}\sim10^{38}$ photons/s to
$\sim10^{42}$ photons/s.
The residual accretion does not change these values much
because the residual accretion rate is never high enough
to significantly increase the mass (and hence the evolution) of the star. 

\begin{figure}
 \centering
 \includegraphics[width= .45\textwidth,angle=0]{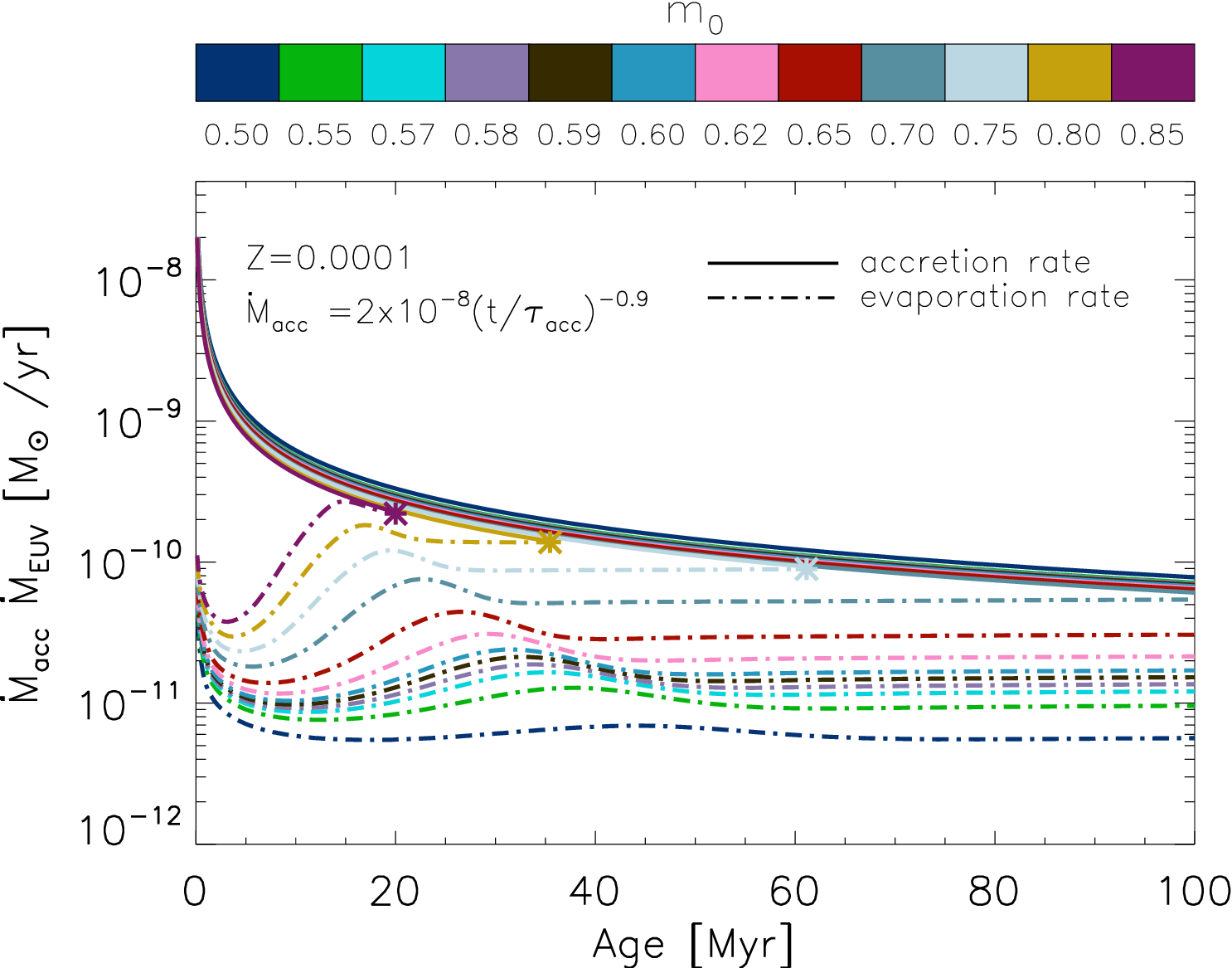}
 \caption{
 Evolution of the residual accretion rate during the first 100 Myr
 along the PMS evolution (solid lines)
 Different colors refer to different initial masses ($m_0$) as shown in the top color-bar.
 The residual accretion begins with the rate
 $2\times10^{-8} ~M_{\odot}/$yr and drops following the Eq.~(\ref{equ:acc}).
  The EUV photo-evaporation (dash-dotted line) rates,  calculated with Eq.~(\ref{equ:euv}) are also plotted.
 From bottom to top are stars with initial mass 0.50$M_{\odot}$ to 0.85 $M_{\odot}$ as labeled in the color-bar.
  The asterisks mark the times when the accretion is terminated by the evaporation.
}
 \label{fig:rate}
\end{figure}

The EUV photo-evaporation rates obtained from Eq.~(\ref{equ:euv})
are shown in Fig.~\ref{fig:rate} (dot-dashed lines)
for a grid of low-mass stars, and they are compared
with the corresponding accretion rates (solid lines) obtained from Eq.~(\ref{equ:acc}).
The EUV photo-evaporation rates increase with increasing initial mass.
Furthermore, at any given initial mass they depend mainly
on the evolution of the EUV luminosity
because the total mass of the star is not significantly affected by the accretion rate.
At relatively larger masses there is an minimum caused by the initial shrinking
 of the radius at almost constant effective temperature (Fig.~\ref{fig:euv}).
Thereafter, the star evolves at constant luminosity toward the ZAMS and, since the effective temperature increases,
the fractional number of EUV photons increases rapidly, so does the evaporation rate.
When the evaporation rate is larger than the accretion rate the accretion effectively stops.
This point is marked with an asterisk in the figure and the terminal age is shown in table~\ref{tab:tstart}.
Beyond this point the star evolves at constant mass.
Figure~\ref{fig:rate} is  meant to provide a schematic diagram
of this mechanism for the first 100~Myr.

  We would like to emphasize that the quenching of accretion is expected to be
  much more complex than our model description.
  For instance, 
we notice that only a fraction of the total EUV photons
from the central star could reach the residual disk, 
if the latter maintains a small geometrical cross section. 
On the other hand, 
stellar activity in young star 
(e.g., accretion shock, magnetic field driven chromosphere activities, etc.), 
which is not included in our model, 
could also be a source of EUV photons and contribute to evaporation. 
This effect could balance the geometrical loss of EUV photons though
the relative contribution between the two sources is not clear.
The end of the late accretion phase could also
depend on the mass of the residual disk before the EUV
evaporation mechanism becomes effective.
For metal-rich star-forming regions, 
disk masses between $0.01-0.2~M_{\odot}$ during the burst accretion phase have been estimated by
~\citet{hartmann1998}. 
While~\citet{bodenheimer} estimate a disk masses from $0.5~M_{\odot}$ down to $0.0001~M_{\odot}$ 
in solar mass YSOs of the Taurus and Ophiuchus star-forming regions.
The residual disk masses we are considering for the present work
are even more uncertain than those of the main disk.
Here we adopt the working hypothesis that
the disk continues to provide  material to the central star until EUV evaporation stops it.

\subsection{A combined PMS model}
A model that combines all the three effects previously discussed is shown in Fig.~\ref{fig:m75}.
The left panel shows the PMS evolutionary track of a star with mass $m_0={0.75}M_\odot$ in the HR diagram.
The dotted line showmass $m_0={0.75}M_\odot$ in the HR diagram.
The dotted line shows the PMS evolution  at constant mass ups the PMS evolution  at constant mass up to deuterium exhaustion.
The latter point represents the end of the stellar birth-line, when the phase of large accretion terminates
and the star begins the contraction at almost constant mass  downward in the HR diagram.
The solid line begins when the deuterium abundance reduces to 1/10 of its initial value.
At this time we apply an accretion with a rate starting at $2\times10^{-8}~M_{\odot}/yr$ (dark blue line in the right panel).
This value is too low to prevent a rapid  contraction
and the central temperature $T_{cen}$ (brown dashed line)  rises until surface Li depletion begins (red line).

\begin{figure}
 \centering
 \includegraphics[width=.45\textwidth,angle=0]{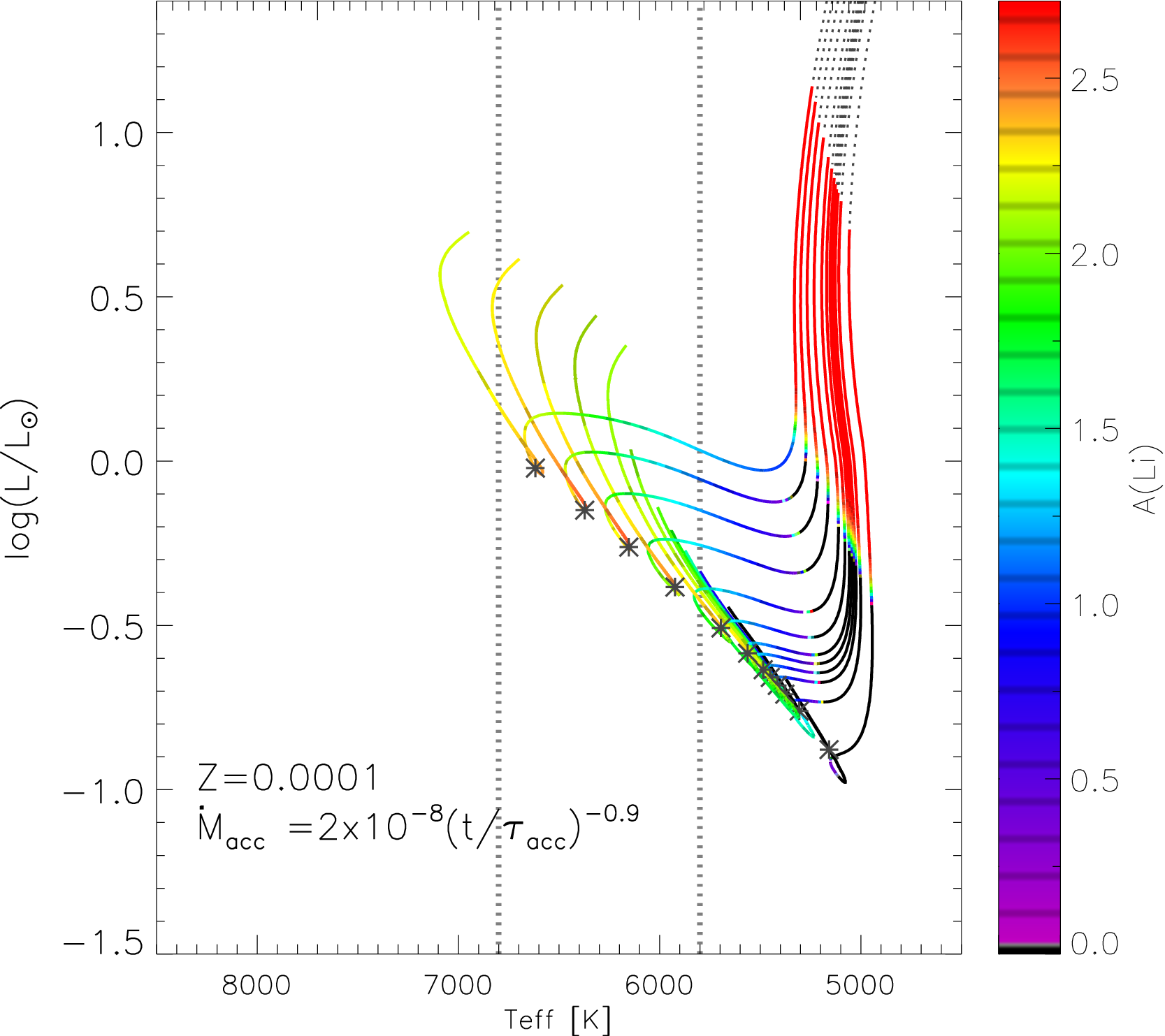}
 \caption{H-R diagram from PMS with overshoot, accretion, and photo-evaporation, to the end of the main sequence.
 The tracks are for stars with initial mass 0.85, 0.80, 0.75, 0.70, 0.65, 0.62,
0.60, 0.59, 0.58, 0.57, 0.55, and 0.50 $M_{\odot}$ from top to bottom.
  Black dashed lines at the upper-right side are Hayashi line with constant initial masses,
 which is not the real case for the late accretion.
 Solid lines are the stellar evolutionary tracks after deuterium burning with an initial lithium value A(Li)=2.72.
Tracks are color-coded according to their surface lithium abundance, as indicated in the right
vertical bar, starting from A(Li)=2.72 (in red) to A(Li) = 0 (violet).
The black color corresponds to A(Li)$<0$.
The area between the two vertical dotted lines is the effective temperature range of the Spite plateau.} 
 \label{fig:hrd}
\end{figure}

At this stage the star is fully convective 
and the surface Li depletion proceeds faster than the restoring effect 
of the accretion because of the high central temperature.
After about five Myr, the core becomes radiative and
the temperature at the base of the convective region $T_{bcz}$ (brown solid line)
begins to decrease followed by a decline of the Li burning rate.
Then the photospheric Li abundance begins to increase again, especially when $T_{bcz}$ falls below $4\times10^6$~K.
This is because on one side Li nuclear burning quenches off and, on the other,
the dilution of the infalling Li becomes weaker as the mass contained in the
the convective envelope becomes smaller.
The mass accreted in this phase is relatively little and  only slightly affects
the total mass of the star with negligible effects on its structure.
In practice, the star evolves at constant mass.

\begin{table*}
 \centering
  \caption{
  Relevant parameters for stars with different initial masses.
  $\tau_{acc}$ is the age at the beginning of the late accretion phase;
  $m_\ast$ is stellar mass in main sequence;
  $t_{end}$ is the age at the end of residual accretion;
  $\Delta n(Li)_{PMS}$ and $\Delta n(Li)_{MS}$
  denote the decrement in  the Li number density
 during pre-main sequence and main sequence, respectively.
  Note that the PMS Li depletion is stronger than the MS one.
  }
  \begin{tabular}{@{}l|r|r|r|r|r|r|r|r|r|@{}}
  \hline
   $m_0~(M_{\odot})$    & 0.57  & 0.58  & 0.59 & 0.60 & 0.62 & 0.65 & 0.70 & 0.75 & 0.80  \\
 \hline
 $\tau_{acc}~(10^5yr)$  & 1.876  & 1.873  & 1.834 & 1.787 & 1.731& 1.689 & 1.595 & 1.560 & 1.433  \\
 $m_\ast~(M_{\odot})$   & 0.616  & 0.625 & 0.633 & 0.641 & 0.658 & 0.685 & 0.729 & 0.775 & 0.821   \\
 $t_{end}~(Myr)$        & 585.55 & 518.96 & 460.83 & 399.76 & 309.84 & 211.46&110.98& 61.14 & 35.46   \\
 $\Delta n(Li)_{PMS}~(10^{-10})$ & 3.16 & 2.81 & 2.60 & 2.50 & 2.33 & 2.11 & 1.75 & 1.17 & 1.21 \\
 $\Delta n(Li)_{MS}~(10^{-10})$  & 0.70 & 0.78 & 0.65 & 0.49 & 0.41 & 0.50 & 0.86 & 1.19 & 1.11 \\
\hline
\end{tabular}
 \label{tab:tstart}
\end{table*}

The surface Li abundance steadily increases towards its primordial value,
and will reach it if accretion does not terminate.
This is due to the shrinking of the surface convective zone which becomes so small that
even a residual accretion rate less than $10^{-10}~M_\odot/$yr is enough to engulf the thin surface convective layers.
The interplay between the size of the convective regions and the evolution of the accretion rates
are critical for reproducing a given pattern in the surface Li abundance as a function of the initial mass on the main sequence.
At an age of about 60 Myr (see Table \ref{tab:tstart}), the residual  accretion is terminated by the EUV photo-evaporation.

The H-R diagram of the full sample of stars considered in this paper is shown
Fig.~\ref{fig:hrd}. The area between the two vertical dotted lines illustrates the temperature range of the Spite plateau.
The high end of this range is also the temperature threshold of the observable lithium.
For stars with higher $T_{\rm eff}$ (the more massive ones), lithium is almost fully ionized in the stellar photosphere.
Likewise in Fig.~\ref{fig:m75}, the upper Hayashi
lines before the completion of deuterium burning are drawn with dotted lines
whilst the solid lines correspond to the stellar evolutionary tracks after deuterium burning.
The surface abundance of Li is color-coded, starting from an initial abundance
A(Li)=2.72.
After the end of the accretion (marked with the asterisks),
stars evolve with constant masses along main sequence and Li abundance no longer increases.

\section{Main sequence Li evolution}\label{sec:ms}

\begin{figure*}
 \centering
 \includegraphics[width= .8\textwidth,angle=0]{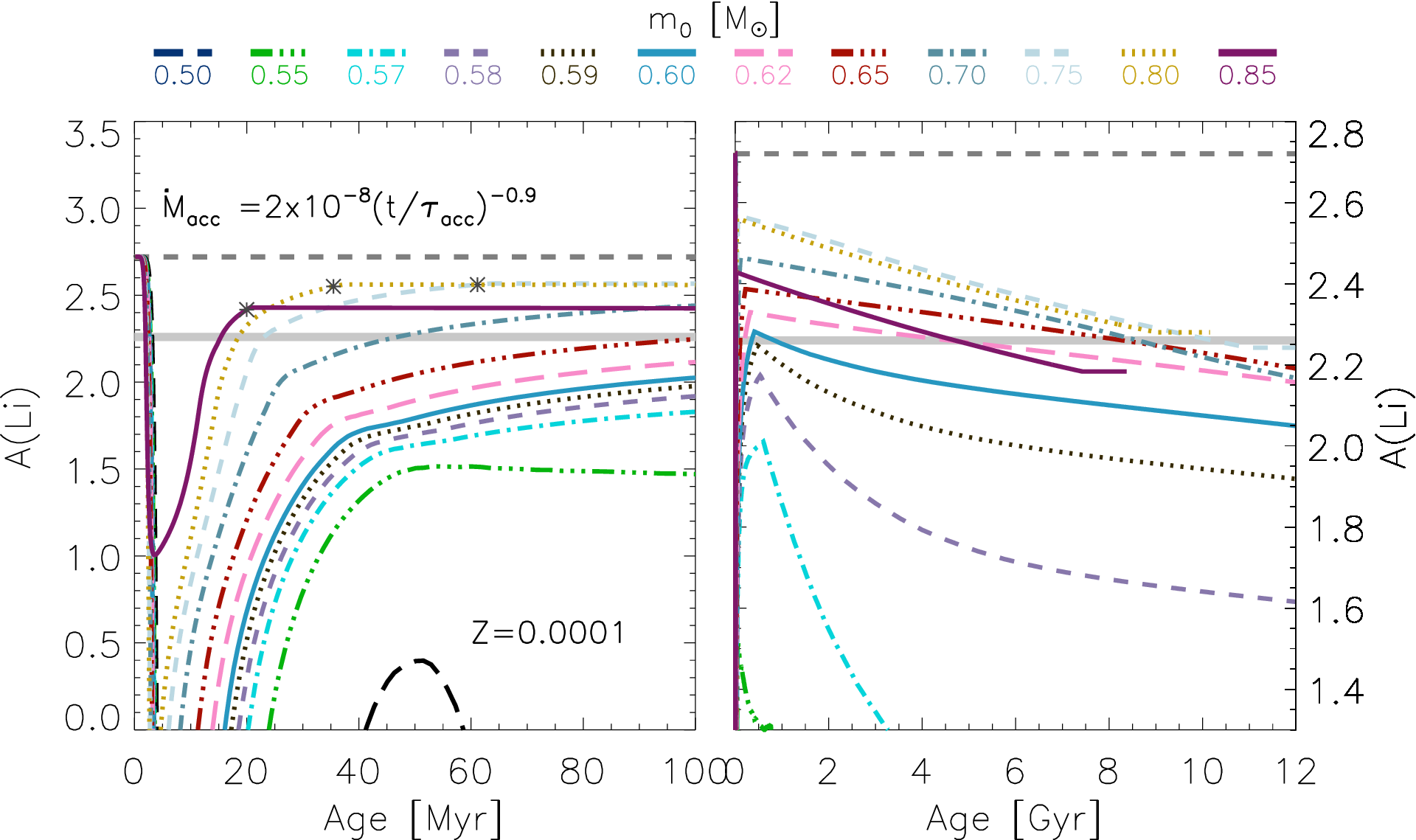}
 \caption{
 A(Li) as a function of the stellar age for stars with different initial masses.
 Different colors refer to  different initial masses ($m_0$).
 The horizontal grey dashed line is the SBBN prediction (A(Li)=2.72), 
 while the horizontal grey solid one indicates A(Li) for the Spite plateau.
 Left panel: Li abundance during the PMS phase during the first 100 Myr.
The initial mass (m$_0$) of the models decreases from 0.85 $M_\odot$ to 0.50 $M_\odot$,  
 from top to bottom in the rapid rising branch on the left of the diagram,
  as indicated in the legend.
 The asterisks mark the end of the accretion.
 Right panel: Li evolution up to an age of 12 Gyr, 
 zooming in around the region around the Spite plateau.
 Color codes and the line styles are the same as in the left panel.
 Note that stars with initial masses greater than $0.80~M_{\odot}$,
 have a main sequence life shorter than 10 Gyr, 
 and they may have already evolved.
}
 \label{fig:ali}
\end{figure*}

During the main sequence phase Li is depleted  by
burning at the base of the convective zone because,
even a low nuclear reaction rate at $T_{bcz}\sim2\times10^6$~K, could cause significant depletion
in a timescale of several Gys. However,
for masses larger than $m_0=0.60M_\odot$, Li burning is insignificant.

Another effect that could  modify the photospheric Li abundance in low mass POP II stars is
microscopic diffusion.
This long-term stellar process cannot be observed directly in the star, and its
efficiency needs to be calibrated.
Microscopic diffusion shortens the main sequence lifetime and leads to a depletion
of the surface elements.
The solar model requires the inclusion of microscopic diffusion,
otherwise even the age of the Sun could not be correctly produced.
We calculated evolutionary  models up to the end of the main sequence including
pressure diffusion, temperature diffusion, and concentration diffusion~\citep{thoul94}.
It has already been shown that microscopic diffusion
without any correction for radiative levitation and surface convective turbulence
cannot reproduce the observed photospheric abundances~\citep{richard02}.
In the case of Li, the surface convective zone of the stars at the plateau high temperature end (the relatively more massive stars) is so thin that,
if microscopic diffusion is at work and not balanced by the radiative levitation,
a too strong depletion is produced.
For example, according to the  model of \citet{sw2001}
we can see that even at $T_{\rm eff}\sim 6200$ K
the predicted Li depletion is too strong compared to the observed plateau.

This does not happen with the standard \texttt{PARSEC} code because
gravitational settling is always inhibited in the outermost region of the envelope, 
$\Delta M=0.5\%$ of the stellar mass \citep{Bressan2013}.
This choice of $\Delta M$ is made according to the suggestion by \citet{chaboyer01}.
They noticed that, while observations indicate that 
the relatively hot low mass stars at the turnoff of the globular cluster NGC 6397 
show the same surface [Fe/H] abundance of the evolved RGB stars \citep{Gratton01},
models with gravitational settling that well reproduce the solar data predict that they should show
a surface [Fe/H] abundance at least 0.28 dex lower. 
They thus conclude that gravitational settling is inhibited in the outermost layers of such stars and show that, 
in order to reconcile the predicted with the observed [Fe/H] abundance, 
the size of this layer should be $\Delta~M\sim~0.5\%~ -~1\% ~M_\odot$ for a star of M$\leq1~M_\odot$.  
We checked that changing the standard \texttt{PARSEC}  parameter
in the range from $\Delta~M= 0.1\%$ to $\Delta~M=1\%$ does not appreciably affect our results.

\section {Results}
\label{sec:res}
We analyse the importance of overshoot and late accretion during the pre-main sequence phase of stellar evolution
by computing a set of evolutionary tracks of initial stellar masses
0.85, 0.80, 0.75, 0.70, 0.65, 0.62,
0.60, 0.59, 0.58, 0.57, 0.55, and 0.50 $M_{\odot}$,
assuming an efficient overshoot, as described in Sect.~\ref{sec:ov},
and exploring different values for the parameters that
describe the form of the residual accretion rate, $\dot{M_0}$  and $\eta$.
We discuss below the results obtained assuming
$\dot{M}_0=2\times{10^{-8}}M_\odot/$yr  and $\eta = 0.9$
in equation \ref{equ:acc},
which produces accretion rates compatible with the observed values.
The models are calculated from the PMS till the end of the main sequence.

We will first focus on the evolution of Li abundance during the PMS phase.
The surface Li abundance evolution of our selected models during the first 100~Myr
of PMS phase is plotted in the left panel of Fig.~\ref{fig:ali}.
With our choice of the parameters, the initial Li evolution in these stars is regulated by overshooting.
The convective overshooting is so efficient that the photospheric Li is significantly
depleted after the first few Myr.
The depletion caused by overshooting gets stronger at the lower masses.
At later times the core becomes radiative and the convective envelope starts receding.

At subsequent stages  Li evolution is governed by the competition between
late accretion, dilution and nuclear burning at the bottom of the convective envelope.
As soon as efficient Li burning ceases, accretion begins to restore the surface Li abundance.
In the more massive stars the convective zone shrinks more rapidly and
Li restoring is faster.
Conversely, in lower mass stars where Li depletion is more efficient
because of the deeper convective zones,
the Li restoring time is longer.

Without EUV evaporation, more massive stars rapidly recover the initial surface
Li abundance while in lower mass stars this process is slower because
nuclear reactions will continue to burn the accreted Li and
the convective dilution is larger.
The effects of EUV evaporation strongly modify this picture and is regulated by the stellar mass.
More massive stars evolve more rapidly toward the ZAMS 
and their EUV luminosity is able to stop accretion at earlier times,
even before the initial abundance is completely restored (see Fig.~\ref{fig:ali}).
On the other hand, in lower mass stars where
Li depletion is more pronounced, the EUV evaporation rates are lower
and accretion, though decreasing, is still able to drive the surface Li abundance
toward the initial value, until it is inhibited by evaporation. 
At even lower masses (e.g., $m_0=0.50M_\odot$),
the recovery of the initial Li  abundance is delayed or even inhibited
by the deep convective zone which is still large when the model reaches the ZAMS.
In this case Li restoring is prevented both by
an efficient dilution, and possibly by the
nuclear burning at the base of such deep convective envelopes.

In summary, our analysis indicates  that Li evolution during the pre-main sequence
phase may be regulated by the competition between
the destroying effect of overshooting at early times,
and the restoring effect of residual accretion,  which is in turn modulated
by EUV photo-evaporation quenching.
This latter effect controls the end result, depending on the mass.
At the higher stellar masses Li restoring is faster,
but accretion quenching takes place earlier.
At lower stellar masses the time-scale for Li restoring is longer,
but the accretion lasts longer because the EUV photo-evaporation has a lower efficiency.
The combination of  these effects in stars with different initial masses
tends to level out the Li abundance just after the PMS at a  value that is already
below the initial one.

\begin{figure}
 \centering
 \includegraphics[width=.46\textwidth,angle=0]{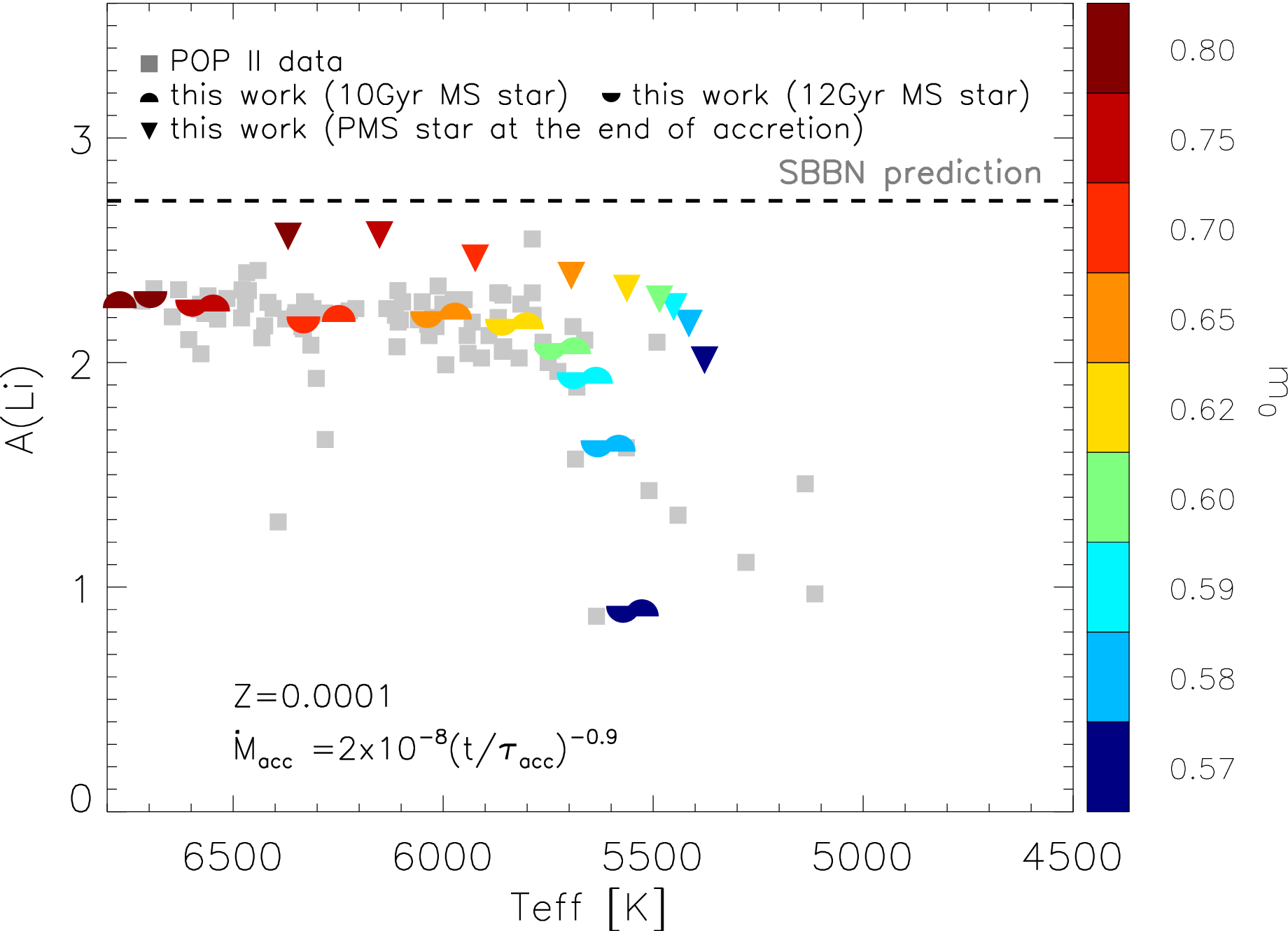}
 \caption{Our results in comparison with the lithium abundance measurements in POP II stars.
 The grey filled squares are POP II data from \citet{molaro12}.
  Our predictions are shown for stars at the end of the late accretion phase (filled triangles), 
  and on the main sequence at 10 Gyr (filled upper circle) and 12 Gyr (filled lower circle).
 Symbols are color-coded according to the initial stellar mass,
 from the left to the right are 0.80 $M_\odot$ to 0.57 $M_\odot$ as the same as labeled.
 The black dashed line marks the primordial Li abundance according to the
  SBBN. }
 \label{fig:com}
\end{figure}

Galactic halo POP II stars we observe today are about 10-12 Gyr old \citep{jw2011}.
In order to reproduce the observed Li abundance of these stars
we have evolved our models up to the end of central hydrogen burning 
(till the turn-off phase).
In this phase we also account for microscopic diffusion
which is known to be a long term effect that
can modify the photospheric element abundances.
The evolution of the photospheric Li abundance during the main sequence
is depicted in the right panel of Fig.~\ref{fig:ali}.
The predicted photospheric Li abundances  are compared with
those of the Galactic halo POP II stars in Fig.~\ref{fig:com}.
We plot the abundances at the end of the accretion phase (filled triangles)
and those at older ages, 10 Gyr (filled upper circles) and 12 Gyr (filled lower circles).
Stars with initial mass $m_0 \le 0.85 M_\odot$ are excluded because
their effective temperatures on the MS are warmer than
the observed ones
(see Fig.~\ref{fig:hrd}), and also
because their main sequence  lifetimes are shorter than the relevant age range.
For our standard choice of parameters,  stars with initial mass
from $m_0 = 0.62 M_{\odot}$ to $0.80 M_{\odot}$,
nicely populate the Spite plateau (A(Li)~$\approx 2.26$).

Models with lower mass, $m_0=0.57 M_\odot$ to $0.60 M_\odot$,
fall on the observed declining branch towards lower temperatures.
We confirm that the first part of this branch is populated by low-mass main sequence stars.
Indeed this is the signature of
the strong Li depletion during the PMS phase (filled triangles),
followed by further depletion during the main sequence evolution (filled upper and lower circles).

Both PMS Li depletion and MS diffusion  contribute to make
the total A(Li) decrease, with the former process playing the main role.
Table~\ref{tab:tstart} lists the Li number density decrease during PMS and MS,
the former is much more than, or at least the same as, the latter,
especially for the lower masses.
The first part of Li declining branch is caused mainly by the PMS depletion.

\begin{figure}
 \centering
 \includegraphics[width=.46\textwidth,angle=0]{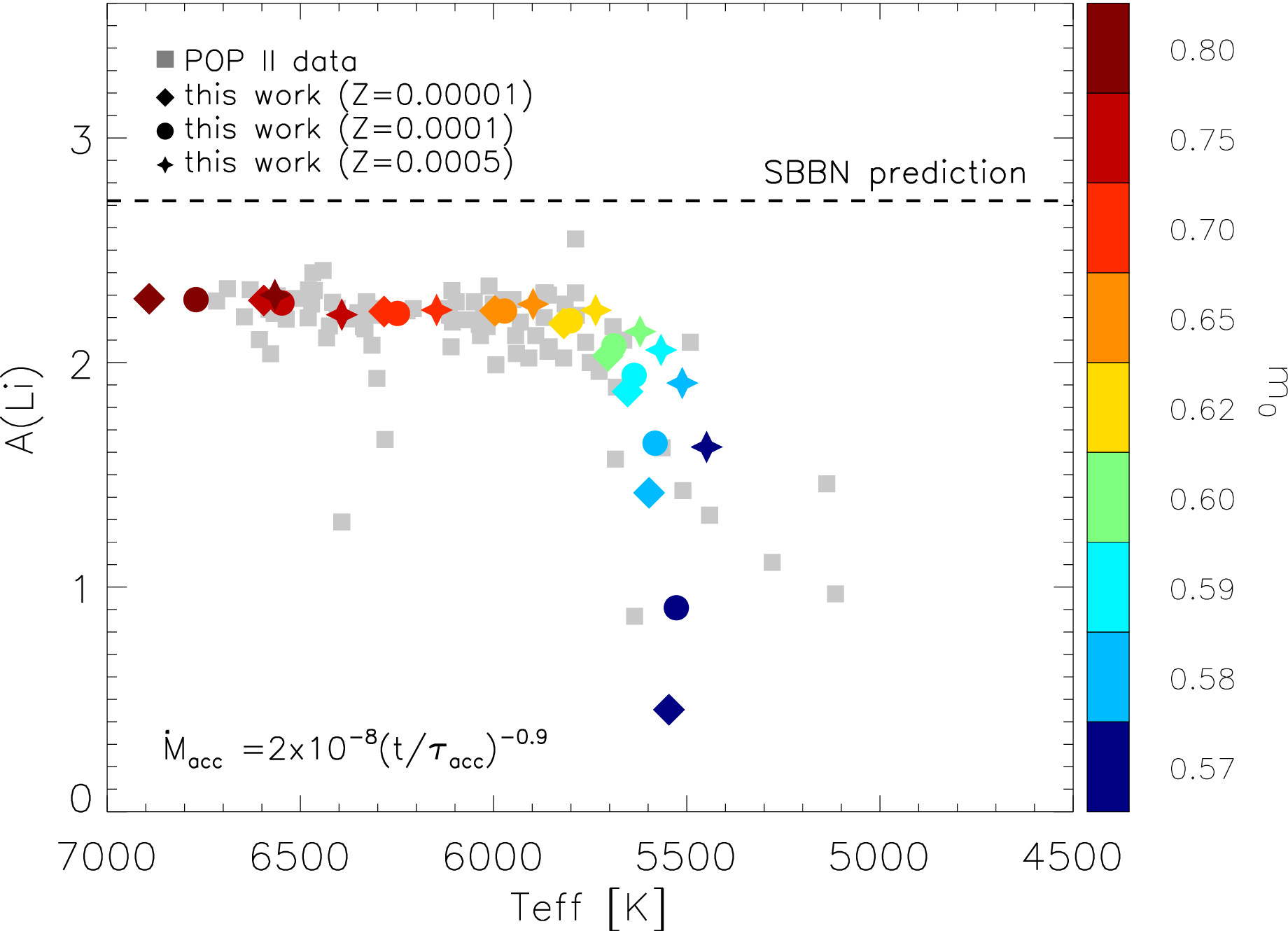}
 \caption{ By applying the same parameter of envelope overshooting and accretion,
 we could reproduce the Spite plateau and the first Li decline branch 
 for a wide range of metallicities (from Z=0.00001 to Z= 0.0005).
 The compared POP II data are the same as in figure~\ref{fig:com}.
 The model results are all main sequence stars at age 10 Gyr, 
 with initial mass 0.80 $M_\odot$ to 0.57 $M_\odot$  from the left to the right for each metallicity as shown in the color-bar label. }
 \label{fig:multi}
\end{figure}

The Spite plateau is populated by metal-poor main sequence stars with different metallicities.
With the same overshooting and accretion parameters we can indeed reproduce the plateau,
including the initial Li declining branch, over a wide range of metallicities,
(from Z=0.00001 to Z=0.0005, that is from  [M/H]=-3.2 to [M/H]=-1.5)
as shown in Fig.~\ref{fig:multi}.

\section{Discussion and Conclusion}\label{sec:dis}
In the previous sections we have shown how the effects of
efficient envelope overshooting, residual accretion
during the pre-main sequence phase, and microscopic diffusion during the main sequence
may modify the photospheric Li abundance in low-metallicity stars.
We have also considered the EUV photo-evaporation process which,
by terminating the accretion phase, could
introduce a sort of self-regulating process.

\begin{figure}
 \centering
 \includegraphics[height=.28 \textheight,angle=0]{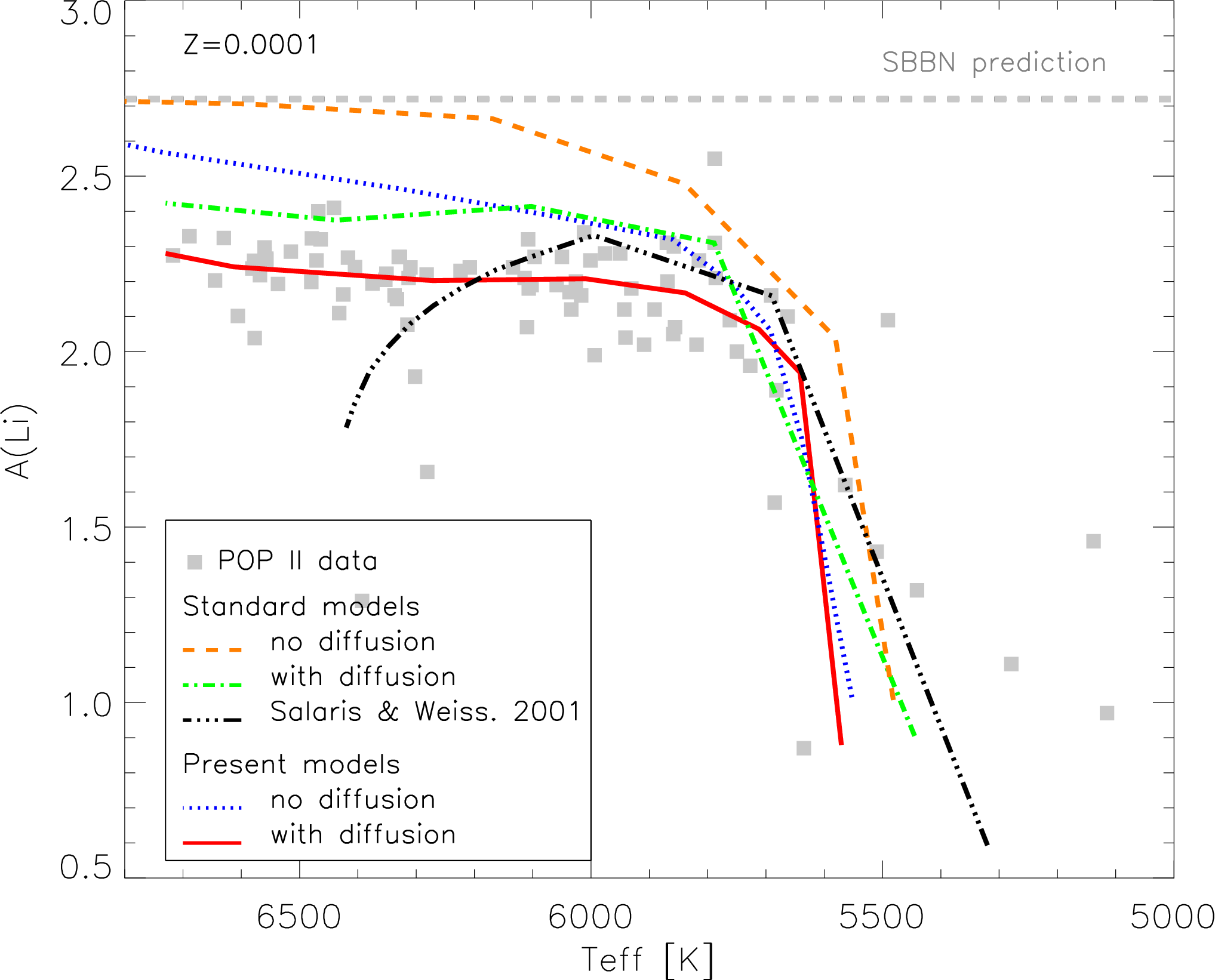}
 \caption{ Comparison of the standard models and the present models for metallicity Z=0.0001
 together with the POP II data and \citet{sw2001} pure diffusion model.
 Different sets of models are classified in the legend.
 The black dash dot dot line is extracted from Fig. 1 of \citet{sw2001} with [Fe/H]=-2.6 and age 12 Gyr.
 }
 \label{fig:models}
\end{figure}

In order to give a more exhaustive picture of the importance of the various processes
we computed more sets of models under different assumptions concerning
the above physical mechanisms.
We synthesize the results of these additional models in Fig.~\ref{fig:models}
where we compare the predicted Li abundances
with the observed ones (see Fig.~\ref{fig:com}).

\begin{itemize}

\item {\bf Standard models }

We begin with two sets of models where none of the PMS effects have been considered.
The first set (orange dashed line) is the standard model without microscopic diffusion while,
in the second set of model (green dot-dashed line) microscopic diffusion
is introduced as it is in the standard \texttt{PARSEC} model.

The model without diffusion is clearly at variance with the observed data.
On the other hand when diffusion is included the Li abundance reaches A(Li)=2.45~dex,
producing a plateau somewhat higher than the observed one, by a factor of 1.6.
The pure diffusion model of \citet{sw2001} is also plotted in this figure
(extracted from figure 1 of \citet{sw2001} with [Fe/H]=-2.6 and age 12 Gyr).
Contrary to them, our diffusive model does not produce the
strong downturn at high effective temperatures.
This is due to the inhibition of diffusion in the outer layers of the star which,
as discussed in Sect.~\ref{sec:ms}, is required to prevent a too strong
sedimentation of helium and heavy elements at the
stellar surface \citep{richard02, parsec}.

While the plateau could eventually be reproduced by increasing the efficiency of diffusion strongly,
we note however that, 
with the adopted diffusion parameters the solar model is very well reproduced \citep{parsec}.

\item  {\bf Present models}

The present  models, that include PMS effects and  diffusion (red solid lines  in Fig.~\ref{fig:models}),
provide a good fit to the  data.
They reproduce both the observed plateau and the declining branch at low temperatures.
As a  test, we recomputed the same  models but switching off microscopic diffusion
during the main sequence phase (blue dotted line).
These new models are identical in the PMS evolution
and incorporate the Li nuclear burning during main sequence.
The latter is mainly responsible for the behavior of Li depletion at low effective temperatures.
They do not reach the plateau value, terminating somewhat at higher Li abundances,
suggesting that some additional depletion is required.
\end{itemize}

In our model we do not  consider stellar rotation.
The additional effects of rotation on Li depletion 
have been discussed in many works, 
but mainly in the context of solar-type stars. 
%Though this paper is devoted to low metallicity stars 
%it is worth summarizing the most recent results.
During the PMS, rotation may affect the internal structure of the stars 
both because of the presence of the centrifugal force and the effects of rotational mixing 
at the base of the convective envelope.
\citet{eggenberger12} conclude that 
by including the full treatment of rotation in PMS,
0.1 dex more Li depletion is obtained.
During the evolution on the main sequence,
rotational mixing tends to oppose to gravitational settling, 
and this solo effect should decrease the surface elements depletion 
with respect to non-rotating models. 
However in the case of Li, 
rotational mixing may be so efficient that 
Li is dragged to encounter temperatures which could efficiently destroy it by nuclear burning.
This is the case of the slow rotator model of M=1.0~$M_\odot$ of \citet{eggenberger10}.
~\citet{garcia94} find that
because of the rotation-caused structural difference, 
stars with faster rotation in open clusters Pleiades and $\alpha$ Persei
have higher Li abundance than the slower rotaters. 
This results is confirmed by 
Pleiades observation of \citet{jones97}.
In general it is found that rotation leads to a large spread 
in the main sequence Li abundance of solar-type stars at varying mass \citep{martin96,mendes99}, 
which could be difficult to reconcile with that observed in the Spite plateau of metal-poor stars.

It is important to stress  that, as long as the stars are accreting at significant rates,
 deuterium burning keep their central temperature around 10$^6$K.
During this phase the stars evolve along the birth line increasing their masses
(e.g. reaching 1 \msun in 0.1 Myr for an accretion rate
of $10^{-5}$ \msun/yr as indicated in table~\ref{tab:tstart}),
while preserving the initial Li abundance \citep{stahler83}.
Once the accretion rates fall below several $10^{-8}$ \msun/yr
the stars abandon the stellar birth-line and evolve at almost constant mass.
It is from this point that we consider the additional effects
of envelope overshooting, which reduces the Li abundance, and of accretion,
 which tends to restore the original abundance.
We have also shown that EUV evaporation of material falling into the stars can act as a natural self-regulating mechanism.

In the relatively more massive stars, where the original Li abundance
could be more quickly restored,
EUV evaporation is more efficient and quenches out accretion earlier than in lower mass stars,
which thus have more time to accrete the pristine gas.
These combined effects tend to level out the Li abundance in stars of different masses.
At even lower masses, Li restoring is inhibited by nuclear burning in the
deep convective envelopes.
After the PMS,  Li is slowly depleted by microscopic diffusion and nuclear burning
which act during main sequence phase.
For stars with $m_0 \geq 0.62 M_\odot$ diffusion drives the main Li depletion whilst,
for $m_0\leq0.60 M_\odot$ stars, Li is significant burned
at the base of the convective envelope.

In this scenario, the observed Spite plateau and its falling branch
at low temperatures are well reproduced by our models,
within a reasonable parametrization of the late accretion mechanism.
Very efficient overshoot is only applied during the early PMS
to  favour strong Li depletion at the beginning of the evolution.
We also find that, even for a maximum envelope overshoot $\Lambda_e = 0.7~H_p$,
which is the value calibrated from the RGB bump \citep{alongi},
the plateau and its falling branch can be finally well reproduced for the same accretion rate.
This indicates that a wide range of envelope overshoot could be applied to the PMS phase.
Even for $\Lambda_e = 0.7~H_p$ the PMS Li depletion is very strong;
without accretion a very low Li abundance would be observed,
e.g. for $m_0=0.70 M_\odot$ A(Li) drops to 1 dex before Li restoring.
It would be important to directly observe whether the very early Li depletion phase exists or not,
because it could confirm that efficient overshoot is at work.
Indeed, at higher metallicities there is evidence of a significant drop of the  Li abundance
during the PMS phase \citep{somers2014}.
On the contrary, the lack of nearby metal-poor star forming regions
prevents any observational test at low metallicity.
This opportunity will be perhaps offered by the next generation telescopes (E-ELT, TMT, and GMT) which could observe the details of low mass metal-poor PMS stars in
star forming regions like those of the Sagittarius dwarf irregular galaxy (metallicity at $Z\sim0.0004$~\citep{sagt}).

For  stars with initial mass $m_0=0.62~M_\odot~-~0.80~M_\odot$, ages in the range 10-12 Gyr, 
and  a wide range of metallicities (Z=0.00001 to Z=0.0005),
our final model is able to predict a present day abundance  consistent with the Spite plateau,
although starting from an initial value of A(Li)=2.72 as inferred from the baryon-to-photon ratio suggested by the CMB
and by the deuterium measurements.
Thus it offers a likely mechanism to solve the long standing
{\it lithium problem}.
It also reproduces the observed lithium drop at the low-temperatures.
The proposed solution relies on stellar physics and evolution,
assuming the validity of the current SBBN theory.
Though this model is quite schematic and rough,
it clearly suggests that the Spite plateau
could be the net result of mechanisms more complex
than those considered so far. In particular, a key role may be
attributed  to the interaction between the stars and their environments.

If the lithium abundance we presently observe in POP II stars   
has been restored in the stellar atmospheres
by a residual tail of accretion after a phase of strong depletion, 
then other puzzling observations could be explained, namely: 
\begin{itemize}
\item{Li at very low metallicities.}

\citet{Sbordone}~found in their observations that for the extreme metal-poor stars,
Li abundance drops when [Fe/H]$<-3$.
~\citet{Caffau11}~and~\citet{Frebel05}~observed dwarfs with [Fe/H] at the lowest levels without any detectable Li.
~\citet{hansen14}~detected Li in an un-evolved star with [Fe/H]=-4.8 at a level of A(Li) = 1.77 and
anther recently discovered extremely metal poor dwarf
SDSS J1742+2531 with [Fe/H] = -4.8 shows A(Li) $<$ 1.8 ~\citep{Bonifacio15}.
The low Li abundance in these stars could be explained by a failed or weaker late accretion.
For the most metal-poor stars, late accretion during the PMS phase might be inhibited and
Li could not be restored.
As we have already shown in Fig.~\ref{fig:pmsov} (right panel),
if no accretion takes place after the initial Li is depleted by the convective overshoot,
Li in these stars would correspond to a very low, or even undetectable, abundance.
Thus the model presented here provides a possible key to account for this  phenomenon, and
we will discuss it in detail in a following paper.

\item{Spectroscopic binaries and outliers}

Our model might also provide an explanation to the Li discrepancy observed in metal-poor binary stars.
The POP II spectroscopic binaries CS 22876-032~\citep{hernandez08} and G 166-45~\citep{aoki2012}
are composed by dwarfs with effective temperatures characteristic of the Spite plateau,
but with the primary stars showing slightly higher Li abundances.
It could simply be an effect of the competition between the
components in the late PMS accretion, with the primary star being the favorite one.

The failure or increase of the accretion process could also provide an explanation
for the few POP II $^7$Li depleted stars or the few POP II stars
that exhibit a $^7$Li abundances at the SBBN prediction level, respectively.
Among the latter there is  BD +23 3912  with  $A(Li)$=2.60
which  stands out from the others in \citet{bm97}.

\item {Lithium behavior in POP I stars}

Li abundance at the solar system formation time,
as obtained from meteorites,  is $A$(Li)$=3.34$ \citep{Anders89}.
The hot F stars of young open clusters never reach this meteoritic value~\citep{ford01}
despite they should have an initial Li value even higher than the  meteoritic  one due to the Galactic Li increase with time.
This Li drop cannot be explained by main sequence Li evolution because the stars are very young.
A PMS Li modification could be the possible mechanism for the depletion.
This will be discussed in detail in a following work dedicated to the metal-rich stars.

\item {Low lithium abundance in the planet host stars}

\citet{Israelian09} and \citet{Gonzalez14} present the observational evidence
that the solar analogues with planet(s)
show lower lithium abundance than those without detected planets.
To explain this effect \citet{eggenberger10}
advance the hypothesis that giant exoplanets prefer 
stars with slow rotation rates on the ZAMS 
due to their longer disk lifetime during the PMS evolution. 
Then their ZAMS models with rotational mixing, which we already discussed in section \ref{sec:ms},
would indicate an enhanced lithium depletion, consistent with observations.
Again it is worth noting that a strong PMS depletion at decreasing mass
is predicted  by solar metallicity models with rotation \citep{mendes99},
and it is not yet clear how this could be reconciled with the observations of 
\citet{Israelian09} and \citet{Gonzalez14}.
In our model, this behavior could be explained if 
somehow the planet(s), located within the accreting disk, 
interfere with the late accretion process.
The PMS accretion could end before being terminated by the EUV evaporation,
inhibiting efficient lithium restoring.
Thus these stars will show lower Li abundance than their analogues without detected planets.
This effect could also explain the larger Li abundance scatter observed in metal-rich stars,
which are more likely to host planets~\citep{lin04}, 
with respect to what observed in metal poor stars.

\end{itemize}

\section*{Acknowledgments}

This research has made use of NASA's Astrophysics Data System.
X.~Fu. thanks Zhi-yu Zhang for helpful discussions. 
A.~Bressan acknowledges PRIN INAF 2014 ``Star formation and evolution in galactic nuclei''.
P. Molaro acknowledges the international team $\sharp272$ lead by C. M. Coppola
``EUROPA- Early Universe: Research On Plasma Astrochemistry'' at ISSI (International Space Science Institute) in Bern,
and the discussions with Giacomo Beccari, Lorenzo Monaco and Piercarlo Bonifacio.
P. Marigo acknowledges support from the University of Padova
({\em Progetto di Ateneo 2012}, ID: CPDA125588/12).

%\bsp

%\label{lastpage}

\end{document}